\documentclass[aps,pra,superscriptaddress,floatfix,showpacs,10pt,showkeys,twocolumn]{revtex4}
\usepackage{amsmath,amsfonts,amssymb,graphics,graphicx,epsfig,color,times,bbm, diagbox,appendix}

\begin{document}

\title{Computable steering criterion for bipartite quantum systems}

\author{Guo-Zhu Pan}
\affiliation{School of Physics {\&} Materials Science, Anhui University, Hefei 230601, China}
\affiliation{School of Electrical and photoelectric Engineering, West Anhui university, Lu'an, 237012, China}

\author{Jun-Long Zhao}
\affiliation{School of Physics {\&} Materials Science, Anhui University, Hefei 230601, China}

\author{Zhi Lin}
\affiliation{School of Physics {\&} Materials Science, Anhui University, Hefei 230601, China}
\affiliation{Department of Physics and State Key Laboratory of Surface Physics, Fudan University, Shanghai 200433, China}

\author{Ming Yang\footnote{mingyang@ahu.edu.cn}}
\affiliation{School of Physics {\&} Materials Science, Anhui University, Hefei 230601, China}

\author{Gang Zhang}
\affiliation{School of Electrical and photoelectric Engineering, West Anhui university, Lu'an, 237012, China}

\author{Zhuo-Liang Cao\footnote{zlcao@ahu.edu.cn}}
\affiliation{School of Physics {\&} Materials Engineering, Hefei Normal University, Hefei 230601, China}
\affiliation{School of Physics {\&} Materials Science, Anhui University, Hefei 230601, China}

\begin{abstract}
Quantum steering describes the ability of one observer to nonlocally affect the other observer's state through local measurements, which represents a new form of quantum nonlocal correlation and has potential applications in quantum information and quantum communication. In this paper, we propose a computable steering criterion that is applicable to bipartite quantum systems of arbitrary dimensions. The criterion can be used to verify a wide range of steerable states directly from a given density matrix without constructing measurement settings. Compared with the existing steering criteria, it is readily computable and testable in experiment, which can also be used to verify entanglement as all steerable quantum states are entangled.
\end{abstract}

\maketitle

\section{Introduction}
It is well known that measurement performed on one part of an entangled quantum state can influence the outcome of the other part without access to it. Such ``spooky action at a distance'' was first noted by Einstein, Podolsky and Rosen in 1935, which aimed to argue the completeness of quantum mechanics \cite{ein}. Later Schr\"{o}dinger introduced the concept of steering in response to the EPR paradox \cite{sch}. In 2007, Wiseman et al. formalized steering in terms of the incompatibility of quantum mechanical predictions with a  classical-quantum model \cite{wis}. Furthermore, the witnessing of quantum steering implies the certification of quantum entanglement without any assumption on one of the parties, i.e., in a one-sided device-independent manner. Steerable states were shown to be advantageous for tasks involving secure quantum teleportation \cite{rei, ros}, quantum secret sharing \cite{walk, kog}, one-sided device-independent quantum key distribution \cite{bra} and channel discrimination \cite{pia}. Because of these, the study of quantum steering has provided new insights to understand quantum theory and consequently has attracted increasing interests recently.

Quantum steering is one form of quantum correlations intermediate between quantum entanglement \cite{horo} and Bell nonlocality \cite{bell}. In the view of quantum information task, quantum steering can be exploited as a resource for quantum communication with one untrusted party, while entangled states need that both parties trust each other and Bell nonlocality is presented on the premise that they distrust each other \cite{jone, brun}. From the view of geometry, quantum states that demonstrate Bell nonlocality lie in a subset of quantum steerable states, while steerable states lie in a subset of entanglement states \cite{wis, qui}. One distinct feature of quantum steering which differs from entanglement and Bell nonlocality is asymmetry. That is, there exists the case when Alice can steer Bob's state but Bob cannot steer Alice's state, which is referred to as one-way steerable and has been demonstrated in theory \cite{bow} and experiment \cite{han, wol}.

Determining whether a quantum state is steerable or not has been one of the fundamental problems in the area of quantum information. Steering inequalities, which are analogous to Bell inequalities, have been proposed to rule out the local hidden variable (LHV)-local hidden state (LHS) models and verify steering \cite{wis}. Recently, a lot of steering inequalities have been derived in discrete and continuous variable systems, such as linear steering inequalities \cite{cav, sau, zhe1, zhe2}, local uncertainty relations steering inequalities \cite{zhen}, covariance matrices steering inequalities \cite{ji}, and entropic steering inequalities \cite{sch2}, etc. Although these steering inequalities work well for a number of quantum states, most of which requires the construction of measurement settings or correlation weights in practice, which increases the complexities of the verification inevitably. It would be desirable to have useful criteria that allow us to verify the quantum steering directly from a given density matrix.

In this paper, we propose a computable steering criterion that is applicable to bipartite quantum systems of arbitrary dimensions. The criterion verifies steering directly from a given density matrix by comparing the values of the purity of the composite system and its subsystem. The purity represents a nonlinear property of a quantum state and can be measured by projecting two copies of the quantum state on symmetric or antisymmetric subspaces \cite{bovi}, so our criterion can be tested directly in experiment. Compared with the existing steering criteria \cite{sau, zhe1, zhe2, zhen, ji, sch2}, ours is more universal as there is no need for us to construct appropriate measurement settings, or select the optimal correlation weights for different types of quantum states. Moreover, our criterion works well for arbitrary-dimensional quantum systems.

\section{Steering criteria for bipartite quantum systems}

Let us first briefly review the steering scenario as introduced by Wiseman et al \cite{wis, jone}. Consider two separated parties, Alice and Bob, sharing an entangled quantum state with density matrix $W$. Alice's task is to convince Bob that the state they shared is entangled. Bob trusts his own apparatus and quantum mechanics, but he does not trust Alice's apparatus, and thus asks her to remotely steer his state by performing local measurements and announce its results through classical communication. The task can be fulfilled if the joint probability distributions cannot be explained by all possible LHV-LHS models in the form
\begin{equation}\label{lhvlhs}
P(a,b|A,B;W)=\sum_{\lambda}P(a|A;\lambda)P(b|B;\rho_{\lambda})p_{\lambda},
\end{equation}
where $P(a,b|A,B;W)$ are joint probabilities for Alice and Bob's
measurements $A$ and $B$, with the results $a$ and $b$,
respectively; $p_{\lambda}$ and $P(a|A;\lambda)$ denote some
probability distributions involving the LHV $\lambda$, and
$P(b|B;\rho_{\lambda})$ denotes the quantum probability of outcome
$b$ given measurement $B$ on the state $\rho_{\lambda}$. In other words,
the state $W$ will be called steerable if it does not satisfy all possible LHV-LHS models. Within this formulation, we propose some
steering criteria that are applicable to bipartite
quantum systems.

In one of our preliminary works, we found a nonlinear steering criterion for two-qubit quantum systems \cite{arxiv_me}, which can be summarized as the following Lemma.

\emph{Lemma 1.} If a given two-qubit quantum state is unsteerable from Alice to Bob (or Bob to Alice), the following inequality holds:
\begin{equation}
\sum\limits_{i=1}\limits^{3}\sum\limits_{j=1}\limits^{3}\langle\sigma_{i}\otimes\sigma_{j}\rangle^{2}\leq1,
\label{lemma}
\end{equation}
where $\sigma_{i,j}$ ($i,j=1,2,3$) are Pauli operators.

Lemma 1 only works for two-qubit quantum systems, so in this paper , we propose a more general steering criterion for bipartite arbitrary-dimensional quantum systems

\emph{Theorem 1.} If a given bipartite quantum state $\rho_{AB}$ is unsteerable from Alice to Bob, the following inequality holds:
\begin{equation}
tr(\rho_{AB}^{2})\leq tr(\rho_{B}^{2}),
\label{computableSC}
\end{equation}
where $\rho_{B}$ is the reduced density matrix for Bob. A brief proof of our theorem is specified below.

\emph{Proof.} For any LHV-LHS model, the joint probabilities of the outcomes $a$ and $b$ upon the measurements $A$ and $B$ for the whole system would satisfy Eq.(\ref{lhvlhs}). Therefore, for a series of measurements $A_{k}$ and $B_{l}$, the observed correlations should satisfy

\begin{small}
\begin{eqnarray}\label{proof}
 \nonumber
&&\sum\limits_{k=1}\limits^{N}\sum\limits_{l=1}\limits^{N}\langle A_{k}\otimes B_{l}\rangle^{2}\\ \nonumber
&=& \sum\limits_{k=1}\limits^{N}\sum\limits_{l=1}\limits^{N}\left(\sum\limits_{a_{k},b_{l}}a_{k}b_{l}P(a_{k},b_{l}|A_{k},B_{l};\rho_{AB})\right)^2\\ \nonumber
&\leq&\sum\limits_{\lambda}\left(p_{\lambda}\sum\limits_{k=1}\limits^{N}\left[\sum\limits_{a_{k}}a_{k}P(a_{k}|A_{k},\lambda)\right]^2\sum\limits_{l=1}\limits^{N}\left[\sum\limits_{b_{l}}b_{l}P(b_{l}|B_{l},\rho_{\lambda})\right]^2\right)\\ \nonumber
&=&\sum\limits_{\lambda}p_{\lambda}\left(\sum\limits_{k=1}\limits^{N}\langle A_{k}\rangle^{2}_{\lambda}\sum\limits_{l=1}\limits^{N}\langle B_{l}\rangle^{2}_{\rho_{\lambda}}\right)\\ \nonumber
&\leq&\kappa C_{A}^{'}\sum\limits_{\lambda}p_{\lambda}\left(\sum\limits_{l=1}\limits^{N}\langle B_{l}\rangle^{2}_{\rho_{\lambda}}\right)\\
&=&\kappa C_{A}^{'}\sum\limits_{l=1}\limits^{N}\langle B_{l}\rangle^{2},
\end{eqnarray}
\end{small}
where $\langle A_{k}\rangle_{\lambda}=\sum_{a_{k}}a_{k}P(a_{k}|A_{k},\lambda)$, $\langle B_{l}\rangle_{\rho_{\lambda}}=\sum_{b_{l}}b_{l}P(b_{l}|B_{l},\rho_{\lambda})$, $C_{A}^{'}=max_{\{\lambda\}}\sum_{k=1}^{N}\langle A_{k}\rangle_{\lambda}^{2}$ and $0\leq\kappa\leq1$. $N$ is the number of the measurement operators for each subsystem. The parameter $\kappa$ is used to adjust the bound to an appropriate value. The first inequality follows from the fact $p_{\lambda}^{2}\leq p_{\lambda}$. The second inequality follows from the definition
of $C_{A}^{'}$. Without loss of generality, we choose an arbitrary complete sets of local orthogonal observables \cite{zhan, yu}, for example
\begin{widetext}
\begin{eqnarray}\label{operators}
A_{k}(B_{l})=\left\{\begin{array}{ccc}
                               &(|m\rangle\langle n|+|n\rangle\langle m|)/\sqrt{2},& \ 1\leq m<n\leq d,\ \ \ for \ \ \ 1\leq k(l)\leq d(d-1)/2,\ \ \ \ \ \ \ \ \ \\
                               &(-i|m\rangle\langle n|+i|n\rangle\langle m|)/\sqrt{2},& \ \ \ 1\leq m<n\leq d, \ \ \ for\ \  d(d-1)/2<k(l)\leq d(d-1),\\                       &|m\rangle\langle m|,& m=1,...,d, \ \ \ \ \ \ \  for\ \ \ d(d-1)<k(l)\leq d^{2},\ \ \ \ \ \ \ \ \ \ \
                             \end{array}\right.
\end{eqnarray}
\end{widetext}
where $d$ is the dimension of the Hilbert space of Alice (or Bob). One has straightforwardly $\sum\limits_{k=1}\limits^{N}\sum\limits_{l=1}\limits^{N}\langle A_{k}\otimes B_{l}\rangle^{2}=tr(\rho_{AB}^{2})$ and $\sum\limits_{l=1}\limits^{N}\langle B_{l}\rangle^{2}= tr(\rho_{B}^{2})$. So the inequality in Eq.(\ref{proof}) reduces to
\begin{equation}\label{purity}
 tr(\rho_{AB}^{2})\leq \kappa'tr(\rho_{B}^{2}).
\end{equation}
where $\kappa'=\kappa C_{A}^{'}$.

For an arbitrary quantum steering criterion, it is preferable to be a sufficient and necessary condition to detect pure states \cite{zhe1, zhe2, zhen}. Here in order to obtain the optimal value of the parameter $\kappa'$, we employ the pure states as reference states. As we know, for any pure separable state $\rho_{AB}$, $tr(\rho_{AB}^{2})=1$ and $tr(\rho_{B}^{2})=1$. $\kappa'$ must satisfy $\kappa'\geq1$ due to the fact that all pure separable states are unsteerable. However, for any pure entangled state $\rho_{AB}$, $tr(\rho_{AB}^{2})=1$ and $tr(\rho_{B}^{2})<1$. Meanwhile $\kappa'$ should satisfy $\kappa'\leq1$ due to the fact that all pure entangled states are steerable. So the optimal value of $\kappa'$ must be $1$. This gives the proof of the theorem 1.

By this way, we derive the steering criterion for arbitrary bipartite quantum systems. Whatever strategies Alice and Bob choose, a violation of
inequality in Eq.(\ref{computableSC}) would imply steering.
\section{Illustrations of generic examples}
In this section, we give some examples of the Theorem 1 applied to some quantum states. By comparing the results with the existing
ones, we show our criterion can verify a wider range of steerable states. For convenience, we call the steering criterion purity
criterion hereafter.

 (i) \emph{Werner state.} The Werner states have been explored extensively in theory and experiment \cite{wern}. For qubits, they can be written as
 \begin{equation}\label{werner}
 \rho_{W}=p|\psi^{+}\rangle\langle\psi^{+}|+(1-p)\mathbb{I}/4,
 \end{equation}
 where $|\psi^{+}\rangle=(1/\sqrt{2})(|00\rangle+|11\rangle)$ is Bell state and $\mathbb{I}$ is the identity, $0\leq p\leq1$. The Werner states are entangled iff $p>1/3$. They are steerable iff $p>1/2$ \cite{wis}. One can get from straightforward calculation that $tr(\rho_{W}^2)=3p^2/4+1/4$ and $tr(\rho_{W})_{B}^{2}=1/2$. According to the purity
criterion, $p>\sqrt{3}/3$ indicates successful steering. Our result is in agreement with the results of Ref. \cite{sau, ji, zhe1, zhe2, zhen}, which implies the steering criterion is qualified for witnessing steering .

\begin{figure}
\centering\includegraphics[width=\columnwidth]{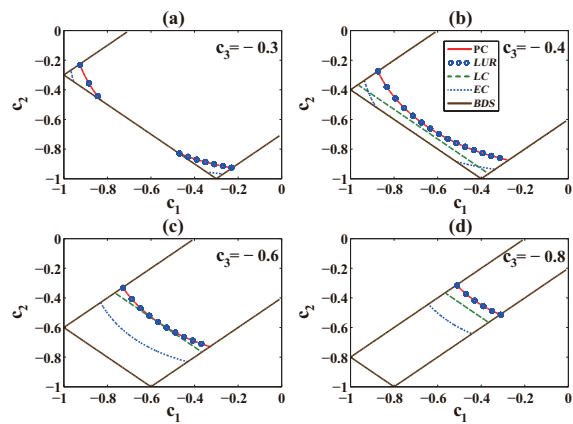}
\caption{The performances of different quantum steering criteria for the Bell diagonal states under the conditions $c_{3}=-0.3, -0.4, -0.6, -0.8$. The area inside the brown solid lines denotes Bell diagonal states (BDS). The red solid lines, blue circled lines,
green dashed lines, blue dotted lines are given by the purity criterion (PC), LUR criterion, linear criterion, entropic criterion,
respectively. States in the left side of these lines are steerable.  It is clear that the PC performs equivalently well as the LUR criterion, which certifies more steerable states than the LC and EC.}
\label{fig1}
\end{figure}

(ii) \emph{Bell diagonal states.} Let us consider the Bell diagonal states shared by Alice and Bob, which can be written as
\begin{equation} \label{bds}
 \rho_{bd}=\frac{1}{4}(\mathbb{I}+\sum_{j=1}^{3}c_{j}\sigma_{j}\otimes\sigma_{j})
\end{equation}
 where $\sigma_{j}$ $(j=1,2,3)$ are Pauli operators and $|c_{j}|\leq1$ for $j=1,2,3$. One can get from straightforward calculation that $tr(\rho_{bd}^{2})=(1+\sum_{j}c_{j}^{2})/4$ and $tr(\rho_{bd})_{B}^{2}=1/2$. Using the purity criterion we find that $\rho_{bd}$ are steerable if $\sum_{j}c_{j}^{2}>1$. Our criterion performs equivalently well as the local uncertainty relations (LUR) steering criterion \cite{zhen}, which certifies more steerable states than the linear criterion (LC) \cite{cav} and entropic criterion (EC) \cite{sch2} (Fig.\ref{fig1}).

(iii) \emph{Asymmetric entangled state.} Consider a asymmetric noisy singlet state of the form
\begin{equation}\label{aes}
\rho_{as}=p|\psi^{-}\rangle\langle\psi^{-}|+(1-p)\rho_{s},
\end{equation}
where $|\psi^{-}\rangle=1/\sqrt{2}(|01\rangle-|10\rangle)$ is a Bell
state and $\rho_{s}=2/3|00\rangle\langle00|+1/3|01\rangle\langle01|$
\cite{guhn}. The state has been demonstrated to be entangled for
$p>0$ by the partial transpose criterion \cite{per}. Using the local
uncertainty relations criterion one can get that it is steerable for
$p>0.536$ in one way and $p>0.582$ in the other way \cite{zhen}.
Another method has conformed it is steerable for $p>0.639$ in one
way and $p>0.604$ in the other way by entropic uncertainty relations
with three mutually unbiased measurements \cite{sch2}. It violates
CHSH-like steering inequality for $p>0.748$ in both ways
\cite{caval}. Using the purity
criterion one finds that $\rho_{as}$ is verified
to be steerable for $p>0.572$ in one way and $p>0.645$ in the other
way. So our method is more powerful than the one in Ref.
\cite{caval} but less powerful than that in Ref. \cite{zhen}.

\begin{figure}
\centering\includegraphics[width=0.8\columnwidth]{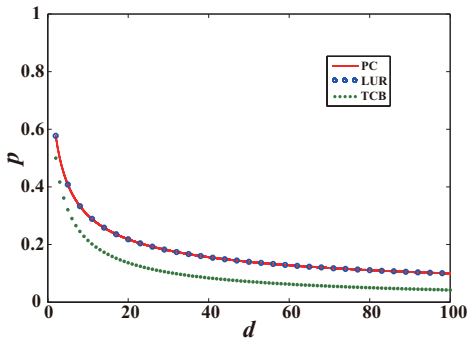}
\caption{The performances of different quantum steering criteria for the isotropic states. The red solid line, blue circled line, and green dotted line are given by the purity criterion (PC), LUR criterion, and theoretical critical bound (TCB), respectively. It is obvious that the PC performs equivalently well as the LUR criterion, which can verify most of the steerable isotropic states.}
\label{fig2}
\end{figure}

(iv) \emph{Isotropic state.} Suppose now that Alice and Bob share a $d\times d$-dimensional isotropic state as follows:
\begin{equation}\label{iso}
\rho_{iso}=p|\varphi_{d}\rangle\langle\varphi_{d}|+(1-p)\mathbb{I}/d,
\end{equation}
where $|\varphi_{d}\rangle=\sum_{i=1}^{d}|ii\rangle/\sqrt{d}$ are maximally entangled states, $0\leq p\leq1$. The isotropic states are entangled iff $p>1/(d+1)$ \cite{horo1}, and steerable iff $p>(\sum_{m=1}^{d}1/m-1)/(d-1)$ in theory \cite{wis}. One has straightforwardly $tr(\rho_{iso}^2)=(d^{2}-1)p^2/d^{2}+1/d^{2}$ and $tr(\rho_{iso})_{B}^{2}=1/d$. By using the purity
criterion, we can obtain $\rho_{iso}$ is steerable when $p>1/\sqrt{d+1}$. In Fig.\ref{fig2}, we plot the area of steerable isotropic states under the LUR criterion, purity criterion and theoretical criterion. It is obvious that the purity criterion can verify most of the steerable isotropic states.

(v) \emph{Free entangled mixed state.} Let us consider a free entangled state of the form \cite{horod}
\begin{equation}\label{free}
\rho_{free}=p|\phi^{+}\rangle\langle\phi^{+}|+(1-p)\sigma^{+},
\end{equation}
where $|\phi^{+}\rangle=1/\sqrt{3}(|00\rangle+|11\rangle+|22\rangle)$, $\sigma^{+}=1/3(|01\rangle\langle01|+12\rangle\langle12|+|20\rangle\langle20|)$ and $0<p<1$. One has straightforwardly $tr(\rho_{free}^2)=4p^{2}/3-2p/3+1/3$ and $tr(\rho_{free})_{B}^{2}=1/3$. According to the purity
criterion, we attain that $\rho_{free}$ is steerable when $p>1/2$. The result is in agreement with the result of Ref. \cite{ji}.

\section{Conclusion}
Being different from the existing steering criteria, our method
verifies the steering directly from a given density matrix without
constructing measurement settings, which implies the steering is an
inherent property of a quantum state. Although the detection of
steering requires us to chooose appropriate measurement settings in
practice, the steerability of a quantum state has nothing to do with
the measurement settings. Our criterion has the following
advantages: (i) The criterion verifies steering only by comparing the
values of the purities of the composite system and its subsystem,
which is readily computable. (ii)As we know, all steerable states
are certainly entangled, so our criterion can be used to verify
quantum entanglement of arbitrary-dimensional bipartite quantum states. In Ref. \cite{wu}, Wu et al. showed that any
quantum state that violate the inequality (\ref{computableSC}) is entangled from the point of
the failure of separable states, which indicates that the criterion is valid
in entanglement verification also. Moreover, for a given steerable
state, the stronger the entanglement is, the higher the violation of the
 inequality (\ref{computableSC}) will be, so our criterion can also be used to quantify entanglement
in some sense. (iii) The criterion can be tested in experiment due
to the successful realization of the direct measurement of the purity \cite{bovi}.

In summary, we have derived a computable steering criterion that is applicable to bipartite quantum systems of arbitrary dimensions. The criterion can be used to verify a wide range of steerable states directly from a given density matrix without constructing measurement settings, which is more universal than the previous ones, and it can be tested in experiment. For a give quantum state, the stronger the entanglement is, the higher the violation of the steering inequality will be, so our criterion can also be used to verify and quantify entanglement in some sense.

\section*{Acknowledgments}

This work is supported by the National Natural Science Foundation of China (NSFC) under Grant Nos. 12004005, 11947102, the Natural Science Foundation of Anhui Province under Grant Nos. 2008085MA16 and 2008085QA26, the Key
Program of West Anhui University under Grant No.WXZR201819,  the Research Fund for high-level talents of West Anhui University under Grant No.WGKQ202001004.


\begin{thebibliography}{99}
\bibitem{ein}A. Einstein, B. Podolsky, and N. Rosen, Phys. Rev. \textbf{47}, 777 (1935).
\bibitem{sch}E. Schr\"{o}dinger, Proc. Cambridge Philos. Soc. \textbf{32}, 446 (1936).
\bibitem{wis}H. M. Wiseman, S. J. Jones, and A. C. Doherty, Phys. Rev. Lett. \textbf{98}, 140402 (2007).
\bibitem{rei}M. D. Reid, Phys. Rev. A \textbf{88}, 062338 (2013).
\bibitem{ros}Q. He, L. Rosales-Z\'{a}rate, G. Adesso, and M. D. Reid, Phys. Rev. Lett. \textbf{115}, 180502 (2015).
\bibitem{walk}N. Walk, S. Hosseini, J. Geng, O. Thearle, J. Y. Haw, S. Armstrong, S. M. Assad, J. Janousek, T. C. Ralph, T. Symul, H. M. Wiseman, and P. K. Lam, Optica \textbf{3}, 634 (2016).
\bibitem{kog} I. Kogias, Y. Xiang, Q. He, and G. Adesso, Phys. Rev. A \textbf{95}, 012315 (2017).
\bibitem{bra} C. Branciard, E. G. Cavalcanti, S. P. Walborn, V. Scarani, and H. M. Wiseman, Phys. Rev. A \textbf{85}, 010301(R) (2012).
\bibitem{pia} M. Piani and J. Watrous, Phys. Rev. Lett. \textbf{114}, 060404 (2015).
\bibitem{horo} R. Horodecki, P. Horodecki, M. Horodecki, and K. Horodecki, Rev. Mod. Phys. \textbf{81}, 865 (2009).
\bibitem{bell} J. S. Bell, Physics \textbf{1}, 195 (1964).
\bibitem{jone} S. J. Jones, H. M. Wiseman, and A. C. Doherty, Phys. Rev. A \textbf{76}, 052116 (2007).
\bibitem{brun} N. Brunner, D. Cavalcanti, S. Pironio, V. Scarant, and S. Wehner, Rev. Mod. Phys. \textbf{86}, 419-478 (2014).
\bibitem{qui} M. T. Quintino, T. V\'{e}rtesi, D. Cavalcanti, R. Augusiak, M. Demianowicz, A. Ac\'{\i}n, and N. Brunner, Phys. Rev. A \textbf{92}, 032107 (2015).
\bibitem{bow} J. Bowles, T. V\'{e}rtesi, M. T. Quintino, and N. Brunner, Phys. Rev. Lett. \textbf{112}, 200402 (2014).
\bibitem{han} V. H\"{a}ndchen, T. Eberle, S. Steinlechner, A. Samblowski, T. Franz, R. F. Werner, and R. Schnabel, Nature Photonics \textbf{6}, 596 (2012).
\bibitem{wol} S. Wollmann, N. Walk, A. J. Bennet, H. M. Wiseman, and G. J. Pryde, Phys. Rev. Lett. \textbf{116}, 160403 (2016).
\bibitem{cav} E. G. Cavalcanti, S. J. Jones, H. M. Wiseman, and M. D. Reid, Phys. Rev. A \textbf{80}, 032112 (2009).
\bibitem{sau} D. J. Saunders, S. J. Jones, H. M. Wiseman, and G. J. Pryde, Nat. Phys. \textbf{6}, 845 (2010).
\bibitem{zhe1} Y. L. Zheng, Y. Z. Zhen, Z. B. Chen, N. L. Liu, K. Chen, and J. W. Pan, Phys. Rev. A \textbf{95}, 012142 (2017).
\bibitem{zhe2} Y. L. Zheng, Y. Z. Zhen, W. F. Cao, L. Li, Z. B. Chen, N. L. Liu, and K. Chen, Phys. Rev. A \textbf{95}, 032128 (2017).
\bibitem{zhen} Y. Z. Zhen, Y. L. Zheng, W. F. Cao, L. Li, Z. B. Chen, N. L. Liu, and K. Chen, Phys. Rev. A \textbf{93}, 012108 (2016).
\bibitem{ji} S. W. Ji, J. Lee, J. Park, and H. Nha, Phys. Rev. A \textbf{92}, 062130 (2015).
\bibitem{sch2} J. Schneeloch, C. J. Broadbent, S. P. Walborn, E. G. Cavalcanti, and J. C. Howell, Phys. Rev. A \textbf{87}, 062103 (2013).
\bibitem{bovi} F. A. Bovino, G. Castagnoli, A. Ekert, P. Horodecki, C. M. Alves, and A. V. Sergienko, Phys. Rev. Lett. \textbf{95}, 240407 (2005).
\bibitem{arxiv_me} G. Z. Pan, M. Yang, H. Yuan, G. Zhang, and J. L. Zhao, arXiv:2010.00083 (2020).
\bibitem{zhan} C. J. Zhang, Y. S. Zhang, S. Zhang, and G. C. Guo, Phys. Rev. A \textbf{76}, 012334 (2007).
\bibitem{yu} S. Yu and N. L. Liu, Phys. Rev. Lett. \textbf{95}, 150504 (2005).
\bibitem{wern} R. F. Werner, Phys. Rev. A \textbf{40}, 4277 (1989).
\bibitem{guhn} O. G\"{u}hne, M. Mechler, G. T\'{o}th, and P. Adam, Phys. Rev. A \textbf{74}, 010301 (R) (2006).
\bibitem{per} A. Peres, Phys. Rev. Lett. \textbf{77}, 1413 (1996).
\bibitem{caval} E. G. Cavalcanti, C. J. Foster, M. Fuwa, and H. Wiseman, J. Opt. Soc. Am. B \textbf{32}, A74 (2015).
\bibitem{horo1} M. Horodecki and P. Horodecki, Phys. Rev. A \textbf{59}, 4206 (1999).
\bibitem{horod} P. Horodecki, M. Horodecki, and R. Horodecki, Phys. Rev. Lett. \textbf{82}, 1056 (1999).
\bibitem{wu} S. J. Wu and J. Anandan, Phys. Lett. A \textbf{297}, 4 (2002).
\end{thebibliography}
\end{document}